\documentclass[twocolumn,showpacs,preprintnumbers,amsmath,amssymb]{revtex4}
\usepackage[dvips]{graphicx}
\usepackage{dcolumn}
\usepackage{bm}

\begin{document}
\title{Equilibrium states and hidden attractors in Chua circuit}
\author{Nataliya~V.~Stankevich$^{1,*}$}
\affiliation{$^{1}$ Yuri Gagarin State Technical University of Saratov, \\
Politechnicheskaya 77, Saratov, 410077, Russian Federation}

\date{\today}
\begin{abstract}
The dynamics of the Chua circuit is studied. Analysis of equilibrium states was revealed. Parameter plane of the picewise linear voltage-current was obtained. For this system was shown sequence of bifurcations of symmetry broken.
$^{*}$Corresponding author. Tel.:+7 9033290994; fax: +7 8452998810

E-mail address: stankevichnv@mail.ru (N.V.Stankevich)
\end{abstract}


\maketitle

\section{Introduction}
Most autonomous, nonlinear dynamic systems that have been studied so far display a structure in which stable periodic, quasiperiodic or chaotic attractors coexist with one (or more) unstable equilibrium points. This is true for the celebrated Lorenz [1] and Rossler [2] oscillators and it is also true for a broad range of other systems that have served as paradigms for the development of nonlinear dynamics. A structure of this form is intuitively acceptable as it conforms to the general understanding that nonlinear dynamic behaviors in autonomous systems develop through destabilization of an original equilibrium state as, with increasing excitation, a cascade of bifurcations leads the system through states of growing complexity.
Recently in Refs. [3-5] was presented studying of the Chua system, where was revealed and visualized hidden chaotic attractor. In [6] the detailed research of that system at the varying of the one parameter was produced. Motivated by the works [3-6]. we re-investigate the hidden attractors reported by [3], and show, that at certain parameters twin hidden attractors can be merged, but its can co-exist separately from each other at another parameters. And we studied basin of attraction of different co-existing attractors.
\subsection{Controlling parameters of the Chua system}
\begin{figure}[h]
\begin{center}
\includegraphics[scale=0.8]{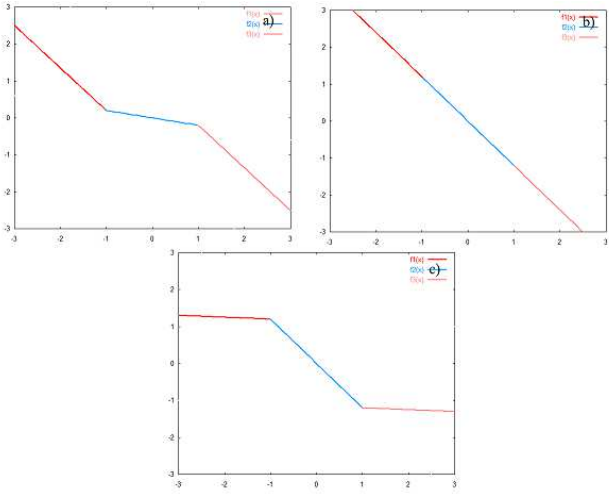}
\end{center}
\caption{Current-voltage characteristic of the system (1) at ?=8.4, ?=12, ?=0.005, a) m0=-0.2, m1=-1.15; b) m0=m1=-1.2; c) m0=-1.2, m1=-0.05.} \label{Fig.1}
\end{figure}

Chua's circuit can be described by the following differential equations:
\begin{equation}
\label{Chua}
  \begin{array}{l l}
    \dot {x}=\alpha(y-x)-\alpha f(x),\\
    \dot {y}=x-y+z,\\
    \dot {z}=-(\beta y+\gamma z),
  \end{array}
\end{equation}
$f(x)=m_{1}x+\frac{1}{2} (m_{0}-m_{1})(|x+1|-|x-1|)$ is piecewise linear voltage-current characteristic. Here $x$, $y$, $z$ are dynamical variables and $\alpha$, $\beta$, $\gamma$, $m_{0}$, $m_{1}$ are parameters of the model. The model (\ref{Chua}) describes dynamics of the electronic circuit []. And parameters $\alpha$, $\beta$, $\gamma$ characterize elements of electronic circuit such resistor, capacitors and inductivity []. Parameters $m_{0}$, $m_{1}$ characterize piecewise linear characteristic of the model, this element introduces nonlinearity to the system. It is well known, that model (\ref{Chua}) has inner symmetry, for replacement $x=-x$, $y=-y$, $z=-z$. This symmetry is provided by symmetry nonlinear characteristic of the model. Parameters $m_{0}$, $m_{1}$ determine slopes of different linear pieces of characteristic, $m_{0}, m_{1}<0$. In dependent on the parameters $m_{0}$, $m_{1}$ one can see the next voltage-current characteristic (Fig.~\ref{Fig.1}). In the case when $m_{0}=m_{1}$, voltage-current characteristic is degenerated to the straight line with slope $m_{0}=m_{1}$.

 In the present paper we consider the dynamics of the system (\ref{Chua}), and co-existing attractors and equilibrium states in dependent on the parameters of piecewise linear characteristic, $m_{0}$ and $m_{1}$. In Sect.2 we provide analytical analysis of the model (\ref{Chua}), and analyze possible equilibrium states, determine its stability in linear approximation. In Sect. 3 we present results of numerical simulations, where we consider opportunity of co-existence of equilibrium states with another hidden attractors. Especial attention will be paid to the question of symmetry attractors, we discuss questions of opportunity of co-existing of symmetry attractors and its mergering.

\section{Equilibrium states of the Chua system}
\subsection{Analytical analysis}
The model (\ref{Chua}) has three equilibrium states: two symmetry points and trivial equilibrium point:
\begin{equation}
\label{Eq_point_1}
\begin{array}{c c}
    x_{0}=\frac{(\gamma+\beta)(m_{0}-m_{1})}{\gamma m_{1}+\beta m_{1}+\beta},
    y_{0}=\frac{(\gamma)(m_{0}-m_{1})}{\gamma m_{1}+\beta m_{1}+\beta},
    z_{0}=\frac{(-\beta)(m_{0}-m_{1})}{\gamma m_{1}+\beta m_{1}+\beta},
\end{array}
\end{equation}
\begin{equation}
\label{Eq_point_2}
\begin{array}{c c}
    x_{0}=0,
    y_{0}=0,
    z_{0}=0,
\end{array}
\end{equation}
\begin{equation}
\label{Eq_point_3}
\begin{array}{c c}
    x_{0}=-\frac{(\gamma+\beta)(m_{0}-m_{1})}{\gamma m_{1}+\beta m_{1}+\beta},
    y_{0}=-\frac{(\gamma)(m_{0}-m_{1})}{\gamma m_{1}+\beta m_{1}+\beta},
    z_{0}=\frac{(\beta)(m_{0}-m_{1})}{\gamma m_{1}+\beta m_{1}+\beta},
\end{array}
\end{equation}

For each equilibrium point we can find analytically condition of Hopf bifurcations. Since voltage-current characteristic is piecewise linear, that in dependent on the position of the point the dynamics near point can be different. For symmetry points matrix of linearization will be the same and has form:
\begin{equation}
\label{Matrix1_3}
    M_{1,3}=\left(
               \begin{array}{ccc}
                 -\alpha (m_{1}+1) & \alpha & 0 \\
                 1 & -1 & 1 \\
                 0 & -\beta & -\gamma \\
               \end{array}
             \right)
\end{equation}

From (\ref{Matrix1_3}) we can get condition for Hopf bifurcation:
\begin{equation}
\label{Hopf_1_3}
    det(M_{1,3})=0 \rightarrow m_{1}=-\frac{\beta}{\beta+\gamma}
\end{equation}

Also we can calculate divegence of the vector field near equilibriums:
\begin{equation}
\label{Div_1_3}
    trace(M_{1,3})=0 \rightarrow m_{1}=-1-\frac{1+\gamma}{\alpha}
\end{equation}
For trivial equilibrium matrix of linearization has little difference::
\begin{equation}
\label{Matrix2}
    M_{2}=\left(
               \begin{array}{ccc}
                 -\alpha (m_{0}+1) & \alpha & 0 \\
                 1 & -1 & 1 \\
                 0 & -\beta & -\gamma \\
               \end{array}
             \right)
\end{equation}
How we can see the difference of matrix of linearization in the parameters $m_{0}$ and $m_{1}$. The matrix (\ref{Matrix1_3}) is independent on the parameter $m_0$, dependent on the parameter $m_1$. And in the matrixes we can realize replacement $m_0=m_1$. It means that bifurcations which happen with symmetry points and trivial point will be the same but for the symmetry points at varying parameter $m_{1}$, and for trivial point at varying parameter $m_{0}$.

So, we get the conditions of Hopf bifurcation for trivial equilibrium:
\begin{equation}
\label{Hopf_2}
    det(M_{2})=0 \rightarrow m_{0}=-\frac{\beta}{\beta+\gamma}
\end{equation}

Also we can calculate divegence of the vector field near equilibrium:
\begin{equation}
\label{Div_2}
    trace(M_{2})=0 \rightarrow m_{0}=-1-\frac{1+\gamma}{\alpha}
\end{equation}

\subsection{Numerical research stability of equilibrium points}
Let us fix parameters of system (\ref{Chua}) near the area, where was obtained hidden attractor in [3], and analyze stability of equilibrium points in dependent on the parameters $m_{0}$ and $m_{1}$. Let  $\alpha=8.4$, $\beta=12$ è $\gamma=0.005$, these parameters are close to the parameters for chaotic hidden attractor which was mentioned in [3]. And let see stability of equilibrium points on the parameter plane ($m_{0}$, $m_{1}$).

\begin{figure}[h]
\begin{center}
\includegraphics[scale=0.8]{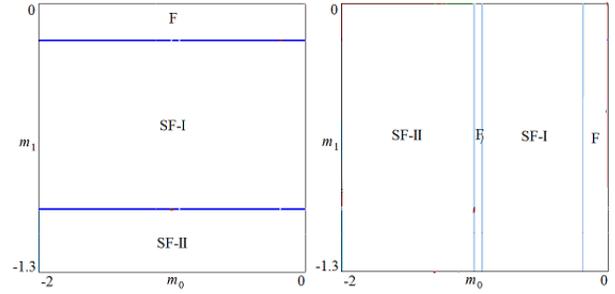}
\end{center}
\caption{Parameter plane of the Chua system (\ref{Chua}), where noticed co-existing equilibrium states.}
\label{Fig.2}
\end{figure}

How was mentioned above, parameters $m_{0}$, $m_{1}$ are characterized slopes of piecewise linear characteristic (see Fig.~\ref{Fig.1}). In the case when $m_{0}=m_{1}$, voltage-current characteristic is degenerated to the straight line (Fig.~\ref{Fig.1}b). How we can see from (\ref{Eq_point_1}), this situation corresponds to the degeneration of equilibrium states to the one: trivial equilibrium state. The replacement of symmetry equilibrium points have place at the intersection of the line $m_{0}=m_{1}$ on the parameter plane.

The model (\ref{Chua}) is three-dimensional, then for this system is possible equilibrium state such saddle-focus. Using matrixes (\ref{Matrix1_3}) and (\ref{Matrix2}) we can get characteristic equations and calculate eigenvalues of the each equilibrium point. For symmetry points eigenvalues will be equal. In. Fig.~\ref{Fig.2} one can see parameter planes, where mentioned lines of changes of signs of eigenvalues for symmetry points (Fig.~\ref{Fig.2}a) and for trivial equilibrium (Fig.~\ref{Fig.2}b).

In Fig.~\ref{Fig.2} we use the next symbols:\\
\textbf{F} is stable focus;\\
\textbf{SF-I} is saddle-focus of the first type: one-dimensional manifold is unstable and two-dimensional manifold is stable;\\
\textbf{SF-II} is saddle-focus of the second type: one-dimensional manifold is stable and two-dimensional manifold is unstable.

So, the dynamical regime corresponding to the chaotic hidden attractor in [3] co-exist with the next equilibrium states: two symmetry \textbf{SF-II} and \textbf{SF-I} in zero.

From Eqs.(\ref{Hopf_1_3}) and (\ref{Hopf_2}) we can get lines corresponding to Hopf bifurcations for $\gamma=12$ and $\beta=0.005$: $m_{0}=-0.9996$, $m_{1}=-0.9996$. This calculations in good agreement with numerical bifurcation lines $l_{m_{1}}^{2}$ and $l_{m_{0}}^{3}$, respectively. Also we put to the plane lines of changing sign of divergency: $l_{m_{1}}^{3s}$ and $l_{m_{0}}^{3s}$.

\section{Numerical studying of parameter plane. Bifurcation scenario of transformation of the hidden attractors}

\begin{figure}[h]
\begin{center}
\includegraphics[scale=0.7]{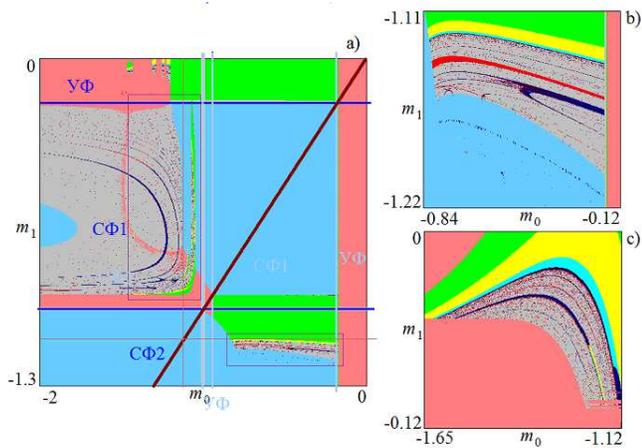}
\end{center}
\caption{Chart of dynamical modes of the Chua system and its magnified fragments, $\alpha=8.4$, $\beta=12$, $\gamma=0.005$, initial conditions: $x_{0}=y_{0}=z_{0}=0.5$.} \label{Fig.3}
\end{figure}

Then let us study the dynamics of the Chua system (\ref{Chua}) numerically. We also will consider the regimes of the system in dependent on the parameter $m_{0}$, $m_{1}$. In Fig.~\ref{Fig.3} the chart of dynamical regimes and its magnified fragment on the parameter plane ($m_{0}$, $m_{1}$) are presented. This charts were constructed in the next way. The parameter plane was scanned with some little step. The dynamical regime was determined after enough long transition process for each point of the parameter plane as amount of discrete points in the Poincar\'{e} section by plane z=0. If in our system there is multistability and co-existence of different attractors and equilibrium states, then we can use different methods of choosing of initial conditions and different directions of scanning of the parameter plane. In the each chart we marked point where we start our calculations and narrow identified direction of scanning. The chart in Fig.3a was constructed with continues method of choosing initial conditions from previous to the next point. We marked stable equilibrium points by red color, the regime of divergency by blue color, chaotic dynamics by gray color (chaotic regime was determined rough: if amount of discrete points in the Poincar\'{e} section was more 120). Another color we use for marking periodic oscillations with different period: green is cycle of period 1, yellow is cycle of period 2, dark-blue is cycle of period 3, blue is cycle of period 4 and etc. Also we have put lines of changing of stability of equilibrium states.How one can see in the Fig.~\ref{Fig.3}3 bifurcations line, getting from the solve of matrixes of linearization and regimes obtained numerically are in a good agreement.

Now let us consider in detail fragment 1, which is shown in Fig.~\ref{Fig.3}b. This area of parameter plane correspond to the chaotic hidden attractor, which was revealed in [1]. In [4] was shown that hidden attractor has twin symmetry attractor, and was considered transformations of attractors at varying of parameter $\alpha$.

In correspondence with analysis of stability of equilibrium points in this area of parameters tree equilibrium point exist: two symmetry saddle-focus \textbf{SF-II} and one (in zero) saddle focus \textbf{SF-I}.  In the fragment 1 one can see line of stabilization of trivial equilibrium state at $m_{0}\approx-0.17$. This line is line of bifurcation of lost stability of zero equilibrium point. In the red area trivial equilibrium point is characterized by one negative real number and two conjugate complex numbers with negative complex part. After cross-section bifurcation line the real parts of  conjugate complex eigenvalues become positive, and stable focus (\textbf{F}) transforms to the saddle-focus with two unstable manifold (\textbf{SF-2}).

How one can see on the chart of the dynamical regimes, at fixed parameter $m_{0}$ and varying of the parameter $m_{1}$, in the system (\ref{Chua}) one can observe transition from limit cycle of period-1 to the chaotic attractor. This transformation realized in corresponding with Feigenbaum scenario: cascade of period-doubling bifurcations.

\begin{figure}[h]
\begin{center}
\includegraphics[scale=0.9]{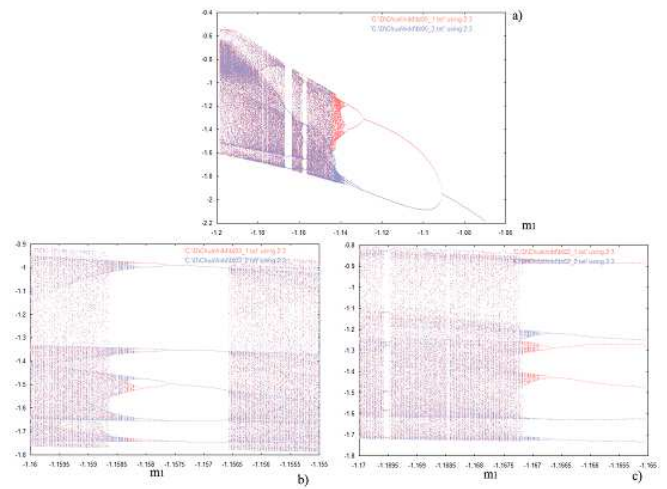}
\end{center}
\caption{Bifurcations diagram: red and violet color correspond to the different initial conditions, $m_{0}=-0.2$.}\label{Fig.4}
\end{figure}

\begin{figure}[h]
\begin{center}
\includegraphics[scale=0.65]{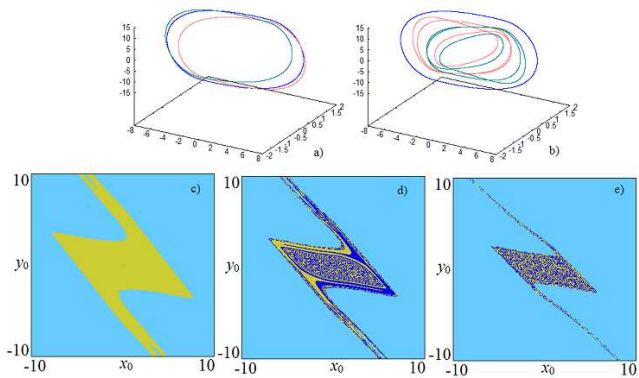}
\end{center}
\caption{a), b) phase portraits; c), d), e) basin of attraction, green and dark blue colors correspond to symmetric attractors.} \label{Fig.5}
\end{figure}

In order to analyze the features of occurring bifurcations, we construct bifurcation diagram in dependence on the parameter $m_{0}$. In Fig.~\ref{Fig.4} the bifurcation diagram and its magnified fragments are shown. How one can see from Fig.~\ref{Fig.4}a, the bifurcation of symmetry broken (or pitchfork bifurcation) at $m_{1}\approx-1.09$ is foregone to the cascade of bifurcations of period doublings. The bifurcation of symmetry broken is typical for Chua model, as this system has inner symmetry. In [4] this bifurcation also was revealed at formation of symmetry twin hidden attractors. Then after bifurcation of symmetry broken on the base of each limit cycle occurs cascade of bifurcations of period doublings. And all attractors co-exist with symmetry twin-attractor. At $m_{1}\approx-1.145$ the twin-attractors are merged in the one. But with further increasing of the parameter $m_{1}$ the bifurcation of symmetry broken take place again, but on base of period-5 cycle (Fig.~\ref{Fig.4}b), corresponding to the window periodicity. Then period -5 cycles undergo cascade of bifurcations of period doublings and chaotic attractors merge again. At further increasing of the parameter m1, we observe co-existence of twin cycles again, but on the base of period-3 cycles and their cascades of period doublings, but in this case there is no previous bifurcation of symmetry broken. In Fig.~\ref{Fig.5} examples of the co-existing twin attractors are shown. In Fig.~\ref{Fig.5} basin of attraction of some co-existing attractors are presented.

\section{Conclusion}
The dynamics of the Chua circuit gives the complex picture in the space of controlling parameters. The sequence of the bifurcations of break symmetry was observed.

This research was leaded in collaboration with N.V.~Kuznetsov, G.A.~Leonov, L.~Chua and supported by the grant of RFBR No.16-32-50012

\end{document}